%% file: sample-sigconf.tex
\documentclass[10pt,conference]{IEEEtran} 
\IEEEoverridecommandlockouts
\usepackage{tikz}
\usepackage{amsmath}
\usepackage{booktabs}
\usepackage{multirow}
\usepackage{listings}
\usepackage{xcolor}
\usepackage{filecontents}
\newcommand{\smallsection}[1]{\noindent {\bf \underline{#1}}.\hspace{1mm}}
\newcommand{\ea}{\textit{et al.}}
\usepackage{url}
\usepackage{xurl}
\usepackage{hyperref}

\usepackage{etoolbox}
\makeatletter
\patchcmd{\@makecaption}
  {\scshape}
  {}
  {}
  {}
\makeatletter
\patchcmd{\@makecaption}
  {\\}
  {.\ }
  {}
  {}
\makeatother
\def\tablename{Table}

\lstdefinelanguage{JavaScript}{
  keywords={typeof, new, true, false, catch, function, return, null, catch, switch, var, if, in, while, do, else, case, break},
  keywordstyle=\color{blue}\bfseries,
  ndkeywords={class, export, boolean, throw, implements, import, this},
  ndkeywordstyle=\color{darkgray}\bfseries,
  identifierstyle=\color{black},
  sensitive=false,
  comment=[l]{//},
  morecomment=[s]{/*}{*/},
  morestring=[b]',
  morestring=[b]"
}

\definecolor{codeblue}{rgb}{0.12, 0.47, 0.71}
\definecolor{codegreen}{rgb}{0,0.6,0}
\definecolor{codegray}{rgb}{0.5,0.5,0.5}
\definecolor{codepurple}{rgb}{0.58,0,0.82}
\definecolor{backcolour}{rgb}{0.95,0.95,0.92}
\definecolor{codered}{rgb}{0.70, 0.09, 0.17}
\lstdefinestyle{mystyle}{
  label=code:sample,
  backgroundcolor=\color{backcolour}, commentstyle=\color{codeblue},
  keywordstyle=\color{magenta},
  numberstyle=\tiny\color{codegray},
  stringstyle=\color{codered},
  basicstyle=\ttfamily\footnotesize,
  breakatwhitespace=false,         
  breaklines=true,                 
  captionpos=t,                    
  keepspaces=true,                 
  numbers=left,                    
  numbersep=2pt,                  
  showspaces=false,                
  showstringspaces=false,
  showtabs=false,                  
  tabsize=1
}
\begin{document}
%-------------------------------------------------------------------------------
% make title bold and 14 pt font (Latex default is non-bold, 16 pt)
\title{Protect Your Secrets: Understanding and Measuring Data Exposure in VSCode Extensions}

\author{
\IEEEauthorblockN{
Yue Liu\IEEEauthorrefmark{2},
Chakkrit Tantithamthavorn\IEEEauthorrefmark{2}\IEEEauthorrefmark{1}, and
Li Li\IEEEauthorrefmark{3}
}
\IEEEauthorblockA{\IEEEauthorrefmark{1}Faculty of Information Technology, Monash University, Melbourne, Australia}
\IEEEauthorblockA{\IEEEauthorrefmark{3}Department of Computer Science and Engineering, Beihang University, China}
Email: \{yue.liu1, chakkrit\}@monash.edu, lilicoding@ieee.org
\thanks{*The Corresponding author.}
}

\maketitle
%-------------------------------------------------------------------------------
\begin{abstract}
Recent years have witnessed the emerging trend of extensions in modern Integrated Development Environments (IDEs) like Visual Studio Code (VSCode) that significantly enhance developer productivity. 
Especially, popular AI coding assistants like GitHub Copilot and Tabnine provide conveniences like automated code completion and debugging. While these extensions offer numerous benefits, they may introduce privacy and security concerns to software developers.
However, there is no existing work that systematically analyzes the security and privacy concerns, including the risks of data exposure in VSCode extensions.

In this paper, we investigate on the security issues of cross-extension interactions in VSCode and shed light on the vulnerabilities caused by data exposure among different extensions. 
Our study uncovers high-impact security flaws that could allow adversaries to stealthily acquire or manipulate credential-related data (e.g., passwords, API keys, access tokens) from other extensions if not properly handled by extension vendors.
To measure their prevalence, we design a novel automated risk detection framework that leverages program analysis and natural language processing techniques to automatically identify potential risks in VSCode extensions.
By applying our tool to 27,261 real-world VSCode extensions, we discover that 8.5\% of them (i.e., 2,325 extensions) are exposed to credential-related data leakage through various vectors, such as commands, user input, and configurations.
Our study sheds light on the security challenges and flaws of the extension-in-IDE paradigm and provides suggestions and recommendations for improving the security of VSCode extensions and mitigating the risks of data exposure.
\end{abstract}
\maketitle
\input{sections/introduction}
\input{sections/background}

\input{sections/challenges}
\input{sections/approach}

\input{sections/experimental_design}
\input{sections/result}

\input{sections/discussion}
\input{sections/related_work}

\input{sections/conclusion}

%-------------------------------------------------------------------------------
% \section*{Acknowledgments}
% %-------------------------------------------------------------------------------

% The USENIX latex style is old and very tired, which is why
% there's no \textbackslash{}acks command for you to use when
% acknowledging. Sorry.

% %-------------------------------------------------------------------------------
% \section*{Availability}
% %-------------------------------------------------------------------------------

% USENIX program committees give extra points to submissions that are
% backed by artifacts that are publicly available. If you made your code
% or data available, it's worth mentioning this fact in a dedicated
% section.

%-------------------------------------------------------------------------------
\bibliographystyle{plain}
\bibliography{sample-base}

\end{document}

%% file: sections/introduction.tex
\section{Introduction}
In recent years, the extension-in-IDE (Integrated Development Environment) has emerged as a predominant trend, revolutionizing the landscape of software development tools.
Among the most prominent platforms is Visual Studio Code (VSCode)~\cite{VSCode_official_page}, a highly versatile IDE developed by Microsoft. 
VSCode operates as a host environment, allowing users to integrate a wide variety of extensions, akin to mini-applications, into the core IDE.
These extensions, ranging from language-specific linters and debuggers to comprehensive project management tools, significantly enhance the functionalities of VSCode.
For example, the popular “GitHub Copilot” extension~\cite{VSCode_Marketplace_copilot} utilizes GPT-4 models~\cite{achiam2023gpt} to generate code suggestions directly within the editor, thereby improving productivity for developers.
Furthermore, the rise of Large Language Models (LLMs) demonstrating exceptional performance in code generation, bug detection, and other software engineering tasks~\cite{liu2022automatically, liu2024refining, liu2024reliability, liu2022explainable} indicates a likely increase in AI-powered extensions enriching the development environment ecosystem.
As of January 2024, the VSCode marketplace hosts over 50,000 extensions, reflecting the remarkable growth and diversity within this ecosystem~\cite{VSCode_Marketplace}. 
This extension-in-IDE paradigm not only addresses the dynamic requirements of software developers, but also encourages a diverse community of extension developers, ranging from individual contributors to large corporations. 

In this extension-in-IDE paradigm, the host VSCode acts like the OS, providing a comprehensive suite of APIs for resource access and interaction, as detailed in the VSCode developer documentation~\cite{VSCodeAPI2023}.
The host VSCode employs a sandboxing mechanism to separate these extensions and their security-critical resources, such as secrets, from each other.
Moreover, VSCode provides interaction interfaces for each extension, such as InputBox and WebView, within the same window. These interfaces are usually inaccessible to other extensions, ensuring the integrity and security of each extension’s user interface.
This mechanism guarantees that extensions operate in isolated contexts, even though they are deeply integrated into the development environment, reducing the risks of cross-extension interference and security vulnerabilities.

\textbf{New security risks in the extension-in-IDE paradigm.} Despite the sandboxing mechanisms employed by VSCode to prevent unauthorized interactions and resource access, unique security challenges emerge in this extension-driven environment. 
Contrasting with mobile apps or the app-in-app models, like the WeChat mini-app system, where different apps typically operate in a distinct and full-screen UI environment, VSCode extensions run in the same workspace.
This shared workspace enables a seamless development experience.
However, extensions can access shared resources like files, configurations, and terminals, potentially leading to security vulnerabilities.
A significant concern may arise when VSCode extension vendors, either unintentionally or through a lack of security practices, expose credentials or critical system functionalities within these shared resources. 
Such exposure can become a target for attackers, who might exploit vulnerabilities in extensions to access or manipulate this information.

To explore these security risks, we conduct a systematic analysis of the data exposure in VSCode extensions.
Our goal is to uncover new attack vectors that enable cross-extension attacks, which can compromise user credentials or other sensitive data.
We identify three main attack vectors that exploit the insecure handling of credential-related data by extension vendors:
(1) accessing secrets stored in a plain-text format in extension configuration or global state; (2) accessing the clipboard to steal credentials users copy from other sources; and (3) binding the credential control with public operations like commands, which allows attackers to control the credentials. 
We demonstrate these risks with concrete examples, even for popular extensions, such as the AI coding assistant \textit{Tabnine v3.62}~\cite{VSCode_Marketplace_tabnine}, which can be exploited to extract the user chat history and send it to the attacker’s servers.
Furthermore, we propose a novel approach to systematically detect and measure the data exposure risks in VSCode extensions. Our approach leverages the advantages of natural language processing (NLP) and program analysis techniques to identify the potential sources and sinks of sensitive data in VSCode extensions. We also classify the types of data that are exposed by the extensions, whether they are credential-related or not, using a fine-tuned BERT model. By applying our approach to a large-scale dataset of real-world VSCode extensions, we uncover a significant number of extensions that are exposed to credential-related data leakage through various vectors, such as commands, user input, and global state. We also analyze the distribution of data exposure risks across different extension categories and popularity levels, revealing interesting patterns and implications for the security of VSCode extensions and the extension-in-IDE paradigm.
We summarize the key contribution below.

\noindent \textbf{• Unveiling Security Risks:} We present the systematic analysis of the risks of data exposure in VSCode extensions. Our study reveals high-impact security flaws that enable adversaries to stealthily acquire or manipulate credentials from other extensions if extension vendors handle them in an insecure or improper manner.

\noindent \textbf{• Automated Detection Framework:} We design and implement a novel automated risk detection framework that leverages the strengths of program analysis and NLP techniques. This framework can extensively and accurately identify potential risks in VSCode extensions, providing a scalable solution for large-scale security analysis.

\noindent \textbf{• Empirical Evaluation:} We apply our framework to a large-scale dataset of over 27,000 real-world VSCode extensions and discover that 8.5\% of them (i.e., 2,325 extensions) are exposed to the credential-related data leakage through various vectors, such as commands, user input, and storage. We also analyze the distribution of data exposure risks across different extension categories and popularity levels, revealing interesting patterns and implications for the security of VSCode extensions and the extension-in-IDE paradigm.

\noindent\textbf{\underline{Open Science.}} To support the open science initiative, we publish the studies dataset and the replication package, which is publicly-available.\footnote{https://github.com/yueyueL/VSCode-Extensions-Security-Analysis}

% \textbf{\underline{Paper Organization.}}
% Section \ref{sec:approach} presents xxx.
% Section~\ref{sec:experiment} presents our studied datasets and the experimental setup, while Section \ref{sec:results} presents our research questions and the experimental results. 
% Section~\ref{sec:relatedwork} presents the related work. 
% Section~\ref{sec:threats} discloses the threats to validity. 
% Section~\ref{sec:conclusion} draws the conclusions. 

%% file: sections/background.tex
\section{Background}
Cross-application security risks on Android applications~\cite{li2015iccta,pan2018flowcog, samhi2022difuzer}, iOS applications~\cite{zhang2018level, xing2015cracking}, web applications~\cite{buyukkayhan2016crossfire}, and mobile in-app applications~\cite{lu2020demystifying, wang2023taintmini, yang2022cross} have been extensively studied in the past.
These research efforts have significantly advanced our understanding of the inherent security challenges and vulnerabilities in these software ecosystems. 
However, little or even no attention has been paid to the IDE platform.
In this section, we briefly introduce the cross-extension isolation on the popular VSCode platform. We then describe how cross-extension attacks on IDEs are more challenging to conduct in practice.

% \begin{figure}[t]
%     \centering
%     \includegraphics[width=0.4\textwidth]{figures/vscode_framework.pdf}
%     \caption{Vscode Extension Architecture}
%     \label{fig:fig_vscode_arch}
% \end{figure}

\subsection{Architecture}
\label{sec:vscode_arch}
VSCode extensions are distributed as packed \textit{.vsix} files.
The \textit{VSIX} archive holds all of the extension files, including the extension's source code (primarily JavaScript/TypeScript), resource files, and a \textit{package.json} file~\cite{VSCodeAPI2023}.
The \textit{package.json} file is a \textit{JSON}-formatted manifest file that provides basic information and configurations about the extension such as its extensionID, version, description, activation events, requested commands, requested configurations, and dependencies.
In addition, a packed \textit{VSIX} file consists of one or multiple JavaScript (JS) files that provide core functionalities of the VSCode extension. It handles commands, interacts with the VSCode API, and manages the extension's lifecycle events.
According to VSCode development documentation~\cite{VSCodeAPI2023}, extension developers utilize either TypeScript or JavaScript for extension development, but TypeScript code undergoes transcompilation into JavaScript prior to the packaging process.

% Unlike Chrome extensions or mini-programs, VSCode extensions have no access to the Document Object Model (DOM) of the VSCode user interface, which means extensions can not directly manipulate or interact with VSCode UI elements through standard DOM methods.
% Despite this, VSCode provides different official APIs for extensions to interact with its user interface and implement their functionality. 
% There are primarily four different approaches: (i) \textit{Command Contributions}, extensions can define commands that can be executed from the command palette, or be bound to keyboard shortcuts, menus, or buttons.
% (ii) \textit{WebView API}, extensions can use the Webview API to create customizable views built with HTML/CSS/JavaScript, which can be displayed next to text editors in the Editor Group areas.
% (iii) \textit{TreeView API}, the Tree View API allows extensions to show content in the sidebar in VSCode. 
% (iv) \textit{Input Box}, the Source Control Input Box, located atop each Source Control view, allows the user to input a message.

\subsection{Security Model}
VSCode employs various security measures to isolate extensions and protect sensitive resources. We summarize the key protections below:

\smallsection{Extension Isolation}
VSCode employs a unique isolation mechanism for its extensions, ensuring that each extension operates within a secure and confined environment.
This isolation is achieved by running each extension in a separate \textit{Node.js} process, known as the extension host.
The extension host is limited to using only the official VSCode APIs, which offer a restricted set of functionalities and interactions with the VSCode platform, such as the user interface or system resources. This restriction ensures that extensions cannot access certain resources. 
For example, extensions cannot directly access the SecretStorage API or the webviews and input boxes belonging to other extensions. This isolation prevents malicious extensions from interfering with the core VS Code application or other extensions.

\smallsection{System Resource Access}
Unlike some popular software systems such as Android or web applications, VSCode does not employ a traditional permission system to regulate API access.
Instead, it provides a series of APIs defined by VSCode itself. These APIs are designed to allow extensions to interact with system resources in a controlled manner.

\smallsection{Extension Vetting}
VSCode employs a rigorous vetting process for extensions published in its Marketplace in order to ensure strong security standards. 
Before publication, every extension undergoes a thorough examination to check code integrity, best practices, and vulnerability absence.
With each new extension submission and update, the VSCode team performs a virus scan on the extension package. This scan serves to certify the extension is free of malware or other security risks.
Through this comprehensive review process, VSCode aims to provide users access only to reliable and high-quality extensions that meet strict security criteria. By adhering to these standards, the VSCode Marketplace fosters an ecosystem of trust between publishers and users. The vetting process provides assurance to users that they can integrate VSCode extensions securely into their development environments without compromising security.

\subsection{Adversary Model}
In our research, we consider an adversary model where the attacker is a malicious extension installed on the VSCode platform. This malicious extension aims to exploit vulnerabilities in other extensions to compromise the confidentiality, integrity, or availability of VSCode, its extensions, or its users.
To clarify, for the attack to be successful, two conditions must be met: (1) A user has installed a malicious extension in a trusted workspace (In this setting, Microsoft’s VSCode security model allows extensions to access the user’s workspace file system and interact with other installed extensions ). (2) The user has also installed a vulnerable extension whose credentials the malicious extension can read or manipulate.
This scenario is not merely theoretical but has been observed in practice.
A recent study~\cite{Checkpoint_Blog_2023} reported that an extension named \textit{prettiest java}, a counterfeit of the popular code formatter \textit{Prettier}, purported to be a Java helper, contained code typical of a Personally Identifiable Information (PII) stealer, seeking local secrets and sending them to the attacker via a Discord webhook.
Since popular extensions often have millions of installs, attackers are incentivized to create malicious extensions masquerading as helpers, add-ons, or even counterfeits of these extensions to exploit their vulnerabilities and large user base.
Such incidents underscore the plausibility of our threat model and the need for robust detection mechanisms.

% % In our research, we studied the potential actions of an isolated extension in collecting sensitive data and utilizing critical resources belonging to other extensions, even when explicitly unauthorized to do so.
% We assume that an adversary can launch cross-app attacks on IDE platforms by exploiting the vulnerabilities of extensions. The adversary’s goal is to compromise the confidentiality, integrity, or availability of VSCode, its extensions, or its users.
% In Section~\ref{sec:attack_mode}, we demonstrate that our adversary can craft a malicious extension that can pass the vetting process of VSCode marketplace and get published.
% In this case, the malicious extension can be installed by users who trust the marketplace and the extension ratings. Once installed, the malicious extension can perform various attacks on the user’s IDE environment, such as stealing credentials or tokens belonging to other vulnerable extensions.

%% file: sections/challenges.tex
\section{Attack Vectors on VSCode}
\label{sec:attack_mode}
In this section, we describe the threat model focusing on potential security risks to user credentials in VSCode extensions.
We identify three aspects of security risks related to user credentials: how extensions store credentials (Section~\ref{sec:risk_storage}), how users input or set credentials (Section~\ref{sec:risk_inputs}), and how extensions control credentials (Section~\ref{sec:risk_control_commands}). 
We show that improper or insecure practices in these aspects can lead to credential leakage, manipulation, or hijacking by malicious extensions. 
To demonstrate the existence of these attacks, we have developed our own proof-of-concept extensions. These extensions were uploaded to the official VSCode marketplace, demonstrating that they can pass the extension vetting process. 
We have provided video evidence of these attacks as further proof of their feasibility. For ethical reasons, we promptly removed these extensions from the marketplace after approval by the vetting process.

\subsection{Improper Credential Storage} 
\label{sec:risk_storage}
Despite the design of VSCode extensions to operate in isolation, not all data within an extension is isolated. For example, configurations and global states can be accessed by other extensions. 
This subsection provides an overview of the storage mechanisms in VSCode extensions and discusses the potential security implications of these mechanisms.

\begin{lstlisting}[language=JavaScript, caption=Example of Data Exposure via Configuration, label=code:configuration_attacker_exp]
// Metadata for vulnerable extensions
// APIkeys are stored in the configuration
"easycode.openAI ApiKey": {
  "type": "string",
  "description": "Your OpenAI Api Key",
}
// Attacker can access the API key
apikey = vscode.workspace.getConfiguration().get('easycode.openAI ApiKey');
\end{lstlisting}

\smallsection{Stealing Data from Extension Configuration}
Configuration is a mechanism that allows VSCode extensions to store and retrieve user preferences and settings.
However, it is not a secure storage, as all the installed extensions can access and modify the current configuration, which can lead to unintended exposure of sensitive data if not properly managed.
Therefore, the extension developers should not store security- or privacy-critical data on configuration. However, some extensions do not follow this best practice and expose personal credentials such as API keys, passwords, or GitHub access tokens in their configuration. This poses a security risk, as an attacker can exploit this vulnerability to steal sensitive data.
For instance, popular AI-driven extensions like \textit{ChatGPT - EasyCode}~\cite{VSCode_Marketplace_EasyCode} (331k installs) store the OpenAI API key in the extension configuration, as shown in Listing \ref{code:configuration_attacker_exp}.
An attacker, through a malicious extension, can easily retrieve this API key using the \textit{vscode.workspace.getConfiguration().get()} method.
This could lead to financial risks as they gain unauthorized access to OpenAI’s ChatGPT models.

\begin{lstlisting}[language=JavaScript, caption=Example of Data Exposure via Global State , label=code:storage_attacker_exp]
// The source code of the vulnerable extension
// The chat history is stored in the global state 
const h = "CHAT_CONVERSATIONS";
const n = e.globalState.get(h, {
    conversations: {}
});
// Update the state with the new conversation
n.conversations[t.id] = {
    id: t.id,
    messages: t.messages
}, await e.globalState.update(h, n)
// Attacker can access the extension storage
// Locate the global state file path
const globalstoragePath = context.globalStorageUri.fsPath;
const directoryPath = path.dirname(globalstoragePath);
const dbFilePath = path.join(directoryPath, "state.vscdb");
// The db file is read and parsed
...
let conversations = valueObject.CHAT_CONVERSATIONS.conversations;
\end{lstlisting}

\smallsection{Stealing Data from Global State}
VSCode allows extensions access to arbitrary files on the host system without explicit user permission. 
This enables malicious extensions to locate and extract sensitive data stored by other extensions. 
One such example is the global state, a memento object that stores key-value pairs that are persisted across VSCode sessions.
The global state is independent of the current opened workspace and can be used to store extension-specific data that does not change frequently.
However, the global state is not isolated from other extensions, as all the installed extensions store their global state values in a file named \textit{“state.vscdb”}, which is a SQLite database.
An attacker can exploit this vulnerability to read, write, or delete the global state of other extensions, potentially compromising their functionality, security, and performance.
For example, an attacker can use the official API, \textit{ExtensionContext.globalStorageUri}, to locate the global state file and read its contents. As shown in Listing~\ref{code:storage_attacker_exp}, an attacker can access the global state of the popular \textit{Tabnine: AI Autocomplete \& Chat} extension~\cite{VSCode_Marketplace_tabnine} (more than 6.8M installs) and extract the chat history. The chat history is stored in the global state under the key \textit{“CHAT\_CONVERSATIONS”}, which contains the conversations between the user and the AI models. The attacker can send this chat history to a third-party server, potentially leaking sensitive information.

\begin{lstlisting}[language=JavaScript, caption=Example of Manipulating Extension Storage, label=code:configuration_update_attacker_exp]
// Metadata for vulnerable extensions
// Webhook used to deliver code is stored in the configuration
"discordCodeShare.webhook": {
    "description": "Webhook used to deliver your code to."
}
// Attacker can update the webhook
const attacker_url = "xxxx"
vscode.workspace.getConfiguration().update('discordCodeShare.webhook', attacker_url);
\end{lstlisting}

\smallsection{Manipulating Extension Storage}
In addition to extracting data, attackers can exploit the lack of isolation in extension storage to manipulate or overwrite an extension’s configuration and states.
This can sabotage functionality or redirect behaviors for malicious objectives without the user’s knowledge.
This can sabotage the extension’s functionality or redirect behaviors for malicious objectives, often without the user’s knowledge. For instance, an attacker can update the configuration using the \textit{vscode.workspace.getConfiguration() .update()} method.
The \textit{tldrdev.discord-code-share} extension~\cite{VSCode_Marketplace_discord-code}, which allows users to send their code to Discord with a few clicks, stores the webhook URL in the configuration. As shown in Listing~\ref{code:configuration_update_attacker_exp}, an attacker can update this URL to point to a malicious webhook. This could lead to unauthorized access to the user’s code, posing a significant security risk.

% \smallsection{Discussion} 
% We implemented three attack extensions on VSCode v1.85.1 to demonstrate the feasibility of these attack vectors. The attack extensions use VSCode official API to steal API keys from the \textit{EasyCodeAI.chatgpt-gpt4-gpt3-vscode v1.2.4} extension, steal the user chat history from the \textit{TabNine.tabnine-vscode v3.62} extension, and send this information to a malicious webhook. Moreover, we manipulate the webhook URL of the \textit{tldrdev.discord-code-share v0.0.2} extension, to redirect the user’s code to our server. The attack extensions successfully passed the vetting of the VSCode marketplace.
% Note that our attack extensions have less than 100 lines of code each, indicating a low barrier to entry for mounting such data exfiltration attacks.
% We reported these issues to the affected extension developers and the VSCode security team. 
% They acknowledged our findings and assured us that they had shared the report with the team responsible for maintaining the product or service. 

\begin{figure}[t]
    \centering
    \includegraphics[width=0.35\textwidth]{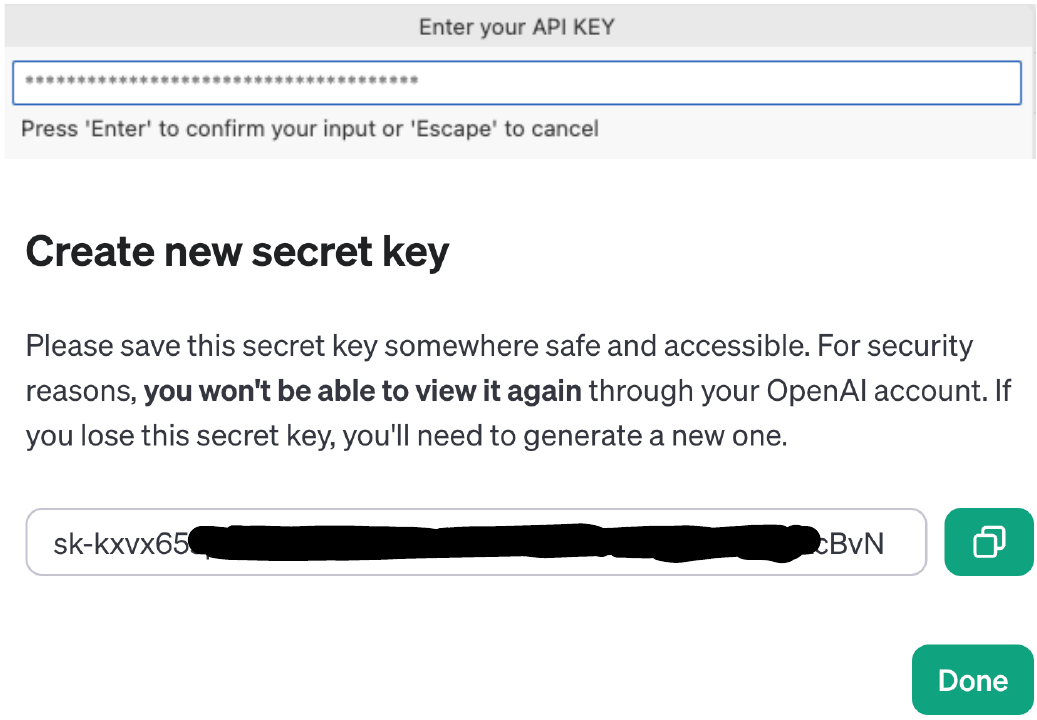}
    \caption{Screenshot for OpenAI APIKey}
    \label{fig:fig_openai_key_exp}
\end{figure}

\subsection{Insecure Credential Input}
\label{sec:risk_inputs}
Clipboard snooping is a security threat that malicious extensions can use to access the clipboard and steal sensitive information that users copy from other sources.
The clipboard is a common way for users to copy and paste content to save time when writing code or configuring their projects, settings, or extensions. 
However, the clipboard is not secure and can be accessed by malicious extensions without the user’s consent or awareness.
To access the clipboard, attacker extensions can use the VSCode API, which provides a method called  \textit{env.clipboard.readText()} that returns the current text content of the clipboard. 
In software development, strings like access tokens and API keys are often long and irregular, which makes them prone to being copied and pasted into the IDE by users. 
A malicious extension could continuously monitor the clipboard to capture sensitive snippets like API keys that users paste.

\begin{lstlisting}[language=JavaScript, caption=Example of Clipboard Access, label=code:clipboard_attacker_exp]
// Source code for potential vulnerable extensions
// Prompting user for sensitive input
window.showInputBox(
{prompt: `${(0,p.translate)("enterYourAPIkey")}
);
// Accessing clipboard contents
const clipboardContent = await vscode.env.clipboard.readText();
\end{lstlisting}

\smallsection{Monitoring the InputBox Clipboard}
VSCode provides an InputBox interface for extension developers to collect user input.
The InputBox interface is a simple dialog box that prompts the user to enter some text, such as a file name, a search query, or a configuration option.
% , as shown in Figure~\ref{fig:fig_openai_key_exp}
Although the InputBox interface is not directly accessible by other extensions, the attacker can still monitor the clipboard for complex but sensitive credentials that the user may copy and paste.
One example of an extension that uses the InputBox interface to collect credentials from users is \textit{zhang-renyang.chat-gpt}~\cite{VSCode_Marketplace_zhang-renyang_gpt}, which is the most popular ChatGPT-related extension in VSCode, with more than 667k installs. 
This extension allows users to chat with various AI models, such as GPT-4 and GPT-3.5, for various purposes, such as code generation. To use this extension, users need to enter their own OpenAI API key, which is a long and complex string that consists of 32 alphanumeric characters and a prefix of “sk-” (see Figure~\ref{fig:fig_openai_key_exp}). 
Users usually copy-paste them from sources like the OpenAI dashboard into the InputBox, rather than typing them by keyboard. 
This behavior enables attackers to continuously monitor the clipboard using \textit{env.clipboard.readText()} and steal any pasted credentials.
As shown in Listing~\ref{code:clipboard_attacker_exp}, the vulnerable extension uses the InputBox to collect the user’s OpenAI key, which the attacker can then access and steal by using \textit{env.clipboard.readText()} when the user copies and pastes the key into the InputBox.

% \smallsection{Discussion} 
% We implemented an attack extension that binds to the \textit{chatgpt.updateAPIkey} command in \textit{zhang-renyang.chat-gpt v1.6.63}. When the target extension prompts the user for their API key via an input box, our extension activates automatically in the background and sends the key to our malicious server via webhook. We uploaded our attack extension to the VSCode marketplace and it passed the vetting process. 

\begin{figure}[t]
    \centering
    \includegraphics[width=0.35\textwidth]{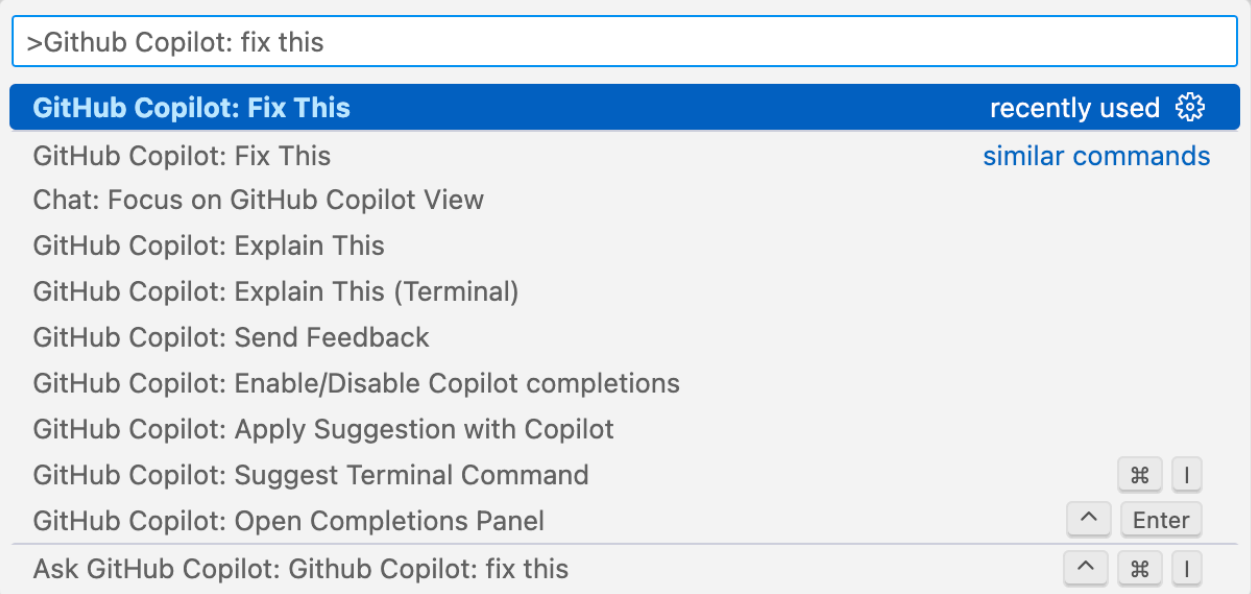}
    \caption{Screenshot for Commands of GitHub Copliot Chat}
    \label{fig:copilot_commands}
\end{figure}

\subsection{Risk Credential Control}
\label{sec:risk_control_commands}
Commands are common UI events in VSCode that allow users to perform various actions, such as formatting code, running tests, or generating content.
As shown in Figure~\ref{fig:copilot_commands}, \textit{GitHub Copilot} provides multiple commands for users to interact with its AI-powered code suggestion features.
Different VSCode extensions can define their own commands and register them with the VSCode API, using unique identifiers such as \textit{“github.copilot.generate”}.
To do so, the extension needs to declare the commands in the \textit{package.json} file and implement the corresponding handlers using the \textit{commands.registerCommand} API in the source code.
Although the identifier is unique, it does not prevent other extensions from invoking these commands either programmatically or manually by the user. This section explores the potential risks associated with the control of user credentials through commands.

\begin{lstlisting}[language=JavaScript, caption=Example of Command Listening, label=code:command_listen]
// Attacker can listen to commands
"activationEvents": [
    "onCommand:github.copilot.generate"
}
\end{lstlisting}

\smallsection{Stealth Command Listening and Binding}
VSCode extensions are activated by certain activation events, such as when a specific language file is opened or a particular command is invoked. 
It means that an extension can listen to the commands of other extensions and activate itself when they are triggered.
This allows a malicious extension to stealthily bind itself to a legitimate command and perform unauthorized actions in the background, such as displaying a fake input box to collect user credentials, sending the captured data to a remote server, or interfering with the expected behavior of the original command.
For instance, we exploited this vulnerability to successfully attack \textit{Github Copilot Chat}~\cite{VSCode_Marketplace_copilot}, a popular AI programming extension developed by the GitHub Team. Our malicious extension listens for the command \textit{“github.copilot.generate”}, as shown in Listing~\ref{code:command_listen}. When this command is invoked, our extension activates and displays an input box, tricking the user into re-entering their Github password to re-login to Github Copilot. Unwitting users, believing they are re-authenticating their session, inadvertently divulge their credentials.

\begin{lstlisting}[language=JavaScript, caption=Example of Manipulating Commands, label=code:command_manipulate]
// Attacker can manipulate commands
vscode.commands.executeCommand('codegpt.removeApiKeyCodeGPT');
const apiKey = await vscode.window.showInputBox({
	  title: 'Enter your API KEY', ...
})
\end{lstlisting}

\smallsection{Manipulating Commands for Secret Exfiltration}
Extensions can define commands to control various operations, including handling sensitive information. Other extensions can execute these operations using the official API \textit{commands.executeCommand}. This can potentially lead to unauthorized access to sensitive information.
For example, the CodeGPT extension~\cite{VSCode_Marketplace_dscodegpt}, another AI programming assistant that supports popular models like OpenAI GPT-4~\cite{achiam2023gpt} and Anthropic Claude~\cite{Anthropic_claude_24}, has a command “codegpt.removeApiKeyCodeGPT” to let the user remove the API key. 
As shown in Listing~\ref{code:command_manipulate}, Attackers can execute this command, prompting the user to re-enter the API key for GPT-4 of OpenAI. The attacker can then automatically display an identical input box to collect the API key.

% \lili{The following one is called typosquatting attacks}
\smallsection{Extension Command Confusion}
Commands are defined based on text strings, which are used as identifiers and labels for the commands.
Although VSCode does not allow extensions to register commands with the same identifier, it does not prevent extensions from using similar or confusing labels for their commands.
An attacker can exploit this vulnerability by creating a command with a label that resembles a command of another extension, such as GitHub Copilot, and registering it with a different identifier. For example, an attacker can register a command with a label \textit{“GitHub CopiIot: Fix This”}, where the lowercase L is replaced by an uppercase i, and use an identifier such as \textit{“attacker.fix”}. As shown in Figure~\ref{fig:copilot_commands}, this label looks almost identical to the original command of GitHub Copilot, which is \textit{“GitHub Copilot: Fix This”}.
This can create ambiguity or deception for the users, who may not be able to distinguish the commands of different extensions or may be misled to run a malicious command instead of a legitimate one.
This may trick the user into running the attacker’s command, which could perform malicious actions such as stealing the user’s code, modifying the code, or redirecting the user to a phishing site.

% \smallsection{Discussion} 
% We implemented two attack extensions to demonstrate the command hijacking vulnerabilities on \textit{GitHub Copilot Chat v0.12} and \textit{DanielSanMedium.dscodegpt v3.1.3}. The attacker extension can bind itself to the existing commands of GitHub Copilot and create confusing commands with similar labels. It can also manipulate the secrets-related commands of CodeGPT to exfiltrate the user’s API key.
% % We reported this issue to the developers of the affected extensions and the VSCode security team.

%% file: sections/approach.tex
\section{Approach}
\label{sec:approach}

\begin{figure*}[t]
    \centering
    \includegraphics[width=0.85\textwidth]{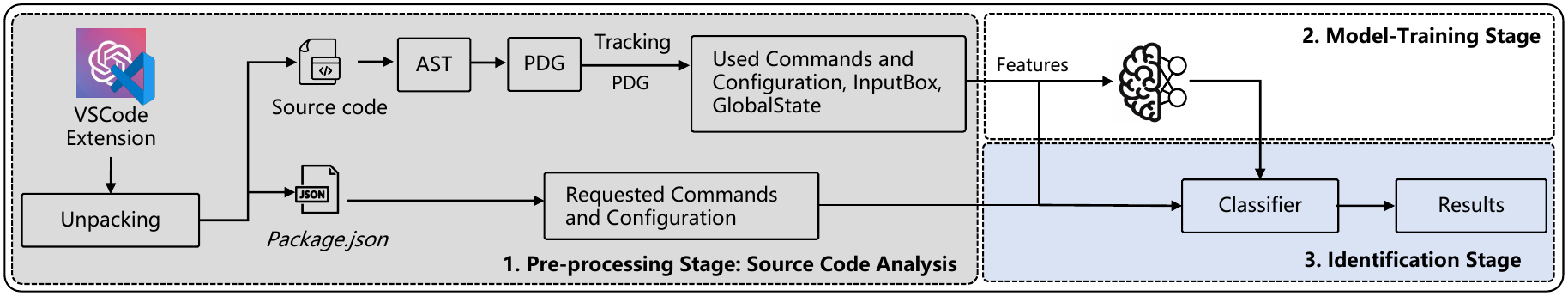}
    \caption{Overview of Our Approach}
    \label{fig:overview_tool}
\end{figure*}

In this section, we describe our automated analysis tool to systematically evaluate the risk of cross-extension attacks in real-world VSCode extensions. Figure~\ref{fig:overview_tool} presents the overview of our approach.

\subsection{Source Code Analysis}
\smallsection{Design}
To analyze the VSCode extensions, we need to extract all the relevant metadata and source code blocks that may contain security risks that described in Section~\ref{sec:attack_mode}. We first unpack each extension into a directory, which includes the \textit{package.json} file and the source code files. 
As described in Section~\ref{sec:vscode_arch}, the \textit{package.json} file is a manifest file that describes the metadata and dependencies of the extension, while the source code files include all the necessary logic and functionality of the extension, such as how it interacts with commands, configuration, and storage.
To comprehend how each extension handles credential-related data (as described in Section~\ref{sec:attack_mode}), we extract the configuration and commands requested in the \textit{package.json} file, and the relevant usage of commands, configuration, input box, and global state in the source code. 
To systematically find all the vulnerable blocks in each VSCode extension, our methodology consists of three steps: (S1) unpacking extensions and extracting metadata like requested commands and requested configuration from the \textit{package.json} file, (S2) building a program dependency graph (PDG) for each source code file and tracking system API calls related to used commands, global state, input box, and used configuration, and (S3) tracking the data flow from the system API calls to the parameters and variables in the PDG.

\smallsection{Step S1: Unpacking Extensions \& Extracting Metadata}
The analysis begins with the unpacking of the VSCode extension’s .vsix file. This file format is a packaging standard for VSCode extensions and contains all necessary files for the extension. The first step involves extracting two critical components: the \textit{package.json} file and the source code files. 
The \textit{package.json} file is essential as it holds vital metadata about the extension, including commands, configuration details, and dependencies. This metadata provides a high-level overview of the extension’s functionality and potential areas for security risks. 
Specifically, we extract the declared \textit{commands} and \textit{configurations} along with their descriptions. This allows us to catalog the requested commands and configurations that the extension can access.
Simultaneously, the source code files are extracted for a detailed analysis of the implementation and internal workings of the extension.

\smallsection{Step S2: Building PDG}
To further enhance our understanding of the source code of an extension, we abstract it with semantic information and model the interactions within and outside of the extension. This approach enables us to perform a data flow analysis to identify potential security risks.
For each extension, we construct an Abstract Syntax Tree (AST) from its source code files. The AST, representing the syntactic structure of the code in a tree format, facilitates programmatic code analysis. From the AST, we enhance it with control and data flows to construct a Program Dependency Graph (PDG). The PDG illustrates how data moves through the code, showing the dependencies and interactions between different variables and functions.
This step is crucial for identifying how the extension interacts with the VSCode API. Part of this process involves locating API nodes within the PDG, which are points in the code where the extension interacts with the VSCode API. These nodes are critical in understanding how the extension uses VSCode’s features and where potential vulnerabilities might exist.

% \lili{Can we present an example of PDG (e.g., by taking one of the code samples in listed in Section 3.1) and demonstrate the data-flow analysis, to help readers better understand}

% \lili{Also, we need to make it clear if we do inter-procedural data-flow analysis or our study only focuses on intra-procedural analysis.}

\input{tables/tab1_api_description}

\smallsection{Step S3: Tracking Data Flow from System API Calls}
In this step, we utilize the PDG to trace data flows from sensitive system API calls to downstream variables and parameters. 
As discussed in Section~\ref{sec:attack_mode}, certain VSCode APIs can expose security risks if they are used improperly for credential management.
To do that, we conduct an intra-procedural data flow analysis.
We identify key API nodes in the PDG corresponding to calls such as accessing configurations, global state, user input, and commands. 
Table~\ref{tab:tab1_api_desc} summarizes the examined API calls that are prone to exploitation described in Section~\ref{sec:attack_mode}.
To systematically pinpoint risks, we execute a depth-first search on the PDG rooted at each identified API node. We recursively traverse data dependencies from API call nodes to extract associated parameter values. This process reveals the specific variables and data dependencies that receive data from sensitive API calls.
Through this rigorous tracking of data provenance, we obtain fine-grained insights into how extensions propagate and store data obtained from privileged VSCode APIs. This approach equips the system to precisely locate open attack vectors that exist due to improper handling of command parameters, user input, configurations, global state, and other sensitive sources.

\subsection{Detecting Data Exposure}

% \lili{Is this step used to identify sources and sinks? If so, I would suggest we make it clear in that way and introduce this section before the code analysis part (which leverages source/sink info for detecting potential security issues)}
% \lili{Normally, when we know the source, we already know the type of data. Why there is a need to further classify them? We need to make that clear.}

As shown in Figure~\ref{fig:overview_tool}, we have extracted the configuration and commands requested in the \textit{package.json} file, along with the API nodes of commands, configuration, input box, and global state from the source code. 
As discussed in Section~\ref{sec:attack_mode}, these elements, if utilized for credential-related purposes, could potentially expose sensitive data. 
Thus, it is important to classify whether a data item is used for credential-related purposes or not. 
To achieve this, we extracted the parameters for each item usage and trained a natural language processing-based classifier to automatically detect the data exposure.

\smallsection{Step D1: Feature Extraction}
In fact, commands, configuration, and input box are types of UI elements that can be used to collect user input or display information. The labels and descriptions of these UI elements can provide semantic clues about the type of data that is being handled by the extension. For example, a label like “Enter your API key” indicates that the extension expects the user to provide a credential. Similarly, the values stored in the global state can also reveal the type of data that is being persisted by the extension. For instance, a key like “CHAT\_CONVERSATIONS” suggests that the extension stores chat history. 
We obtain these textual features from two sources: the metadata and the PDG. From the metadata, we extract the \textit{requested commands} and \textit{requested configurations} along with their descriptions. From the PDG, we extract the \textit{used commands}, \textit{used configuration}, \textit{global state}, and \textit{inputbox} along with their labels and parameters. These features form the input for our data classification model.

\smallsection{Step D2: Training Data}
To train our data classification model, we need a labeled dataset of VSCode extensions and their exposed data types. However, such a dataset does not exist, so we manually create one by sampling 500 extensions from the official marketplace (see Section~\ref{sec:experiment}). We inspect the source code and metadata of each extension and label them with one of the following two categories: (1) credentials-related, such as API keys, passwords, access tokens, etc., (2) others that are not sensitive, such as code snippets, configuration options, etc. We use the text features extracted from the UI elements and global state keys as the input for each extension, and the corresponding label as the output.

\smallsection{Step D3: Text Classifcation}
Finally, we use a text classification model to learn the semantic patterns and associations between the text features and the data types. We use a pre-trained BERT model~\cite{kenton2019bert} as the backbone of our classifier, as it has shown state-of-the-art performance on various natural language understanding tasks. We fine-tune the BERT model on our labeled dataset, using a linear layer on top of the BERT output to predict the data type label.

%% file: tables/tab1_api_description.tex
% Table generated by Excel2LaTeX from sheet 'sinks_sources'
\begin{table}[t]
  \centering
  \footnotesize
  \caption{Summary of VSCode API Calls}
  \scalebox{0.85}{
    \begin{tabular}{lll}
    \toprule
    Name  & \multicolumn{1}{c}{API} & \multicolumn{1}{c}{Description} \\
    \midrule
    \multirow{2}[1]{*}{Commands} & \textit{registerCommand} & Invokes a command \\
          & \textit{registerTextEditorCommand} & Invokes a text editor command \\
    Configuration & \textit{workplace.getConfiguration} & Obtains a configuration value \\
    InputBox & \textit{window.showInputBox} & Collects user input \\
    GlobalState & \textit{globalState.update} & Updates a key in the global state \\
    \bottomrule
    \end{tabular}%
    }
  \label{tab:tab1_api_desc}%
\end{table}%

%% file: sections/experimental_design.tex
\section{Experimental Design}
\label{sec:experiment}

\smallsection{VSCode Extensions Collection (denoted as $D_t$)}
For our research, we collected a comprehensive dataset of VSCode extensions from the official VSCode marketplace.  We crawled the marketplace website in September 2023 and downloaded the latest versions of all available extensions, resulting in a total of 48,692 extensions. These extensions span 18 categories, such as themes, data science, languages, and AI, reflecting the rich and varied functionality of VSCode.
To focus our analysis on the most relevant and popular extensions, we applied two filtering criteria: (1) we excluded extensions that had fewer than 10 installs, and (2) we excluded extensions that did not contain any source code files. After applying these filters, we obtained a final dataset of 27,261 extensions, which we used for our subsequent analysis.
The extensions in our dataset have an average of 70k installs, with the most installed extension being \textit{Python} with almost 100 million installs. The average size of the extensions is 3.8 MB, ranging from 1.5 KB to 567 MB. This extensive and varied dataset provides a robust foundation for our security analysis of VSCode extensions.

\smallsection{Ground-truth datasets (Denoted as $D_{gt}$)}
In order to establish a ground-truth dataset, we randomly selected 500 extensions from our initial collection $D_t$. These extensions were then manually labeled by a researcher and a developer, both of whom have extensive experience with VSCode. They independently identified and labeled data blocks extracted through source code analysis.
The inter-rater agreement was measured using Cohen’s kappa coefficient, which was 0.95 for data points, indicating a high level of consistency.
Any disagreements between the two labelers were resolved through discussion.
From the 500 extensions, we identified 444 data points that are credential-related and 16,512 data points that are normal.
We used this ground-truth dataset to evaluate the effectiveness and accuracy of our automated analysis tool.

\smallsection{Implementation}
We implemented the prototype of our approach using Javascript and Python scripts. For the source code analysis, we used Espree~\cite{Espree2023} to build AST for VSCode extensions, and extended DoubleX~\cite{fass2021doublex} to build PDG and perform data flow analysis. We conducted all experiments on a 32-core 64GB of memory Linux 5.15.0 kernel machine.
To train and evaluate the text classification model, we used the ground truth dataset $D_{gt}$.
We implemented the BERT-based classifier using PyTorch~\cite{paszke2019pytorch} and transformers~\cite{wolf2019huggingface} libraries. 
We used cross-entropy loss for training the model. To handle class imbalance in the dataset, we adjusted the loss function by assigning different weights to different classes.

%% file: sections/result.tex
\section{Evaluation}
\label{sec:results} 

\subsection{Effectiveness}
To evaluate the effectiveness of our approach, we first performed a rigorous 10-fold cross validation on our ground truth dataset $D_{gt}$ consisting of 16,958 labeled data points across 500 extensions.
The 10-fold cross-validation was designed to predict unseen extensions, providing a robust assessment of our model’s generalizability.
For the classifier, we fine-tuned a BERT model on $D_{gt}$ to predict whether a data point is credential-related or normal. 
We used a softmax output layer and a cross-entropy loss function with class weights of 0.01 for credential-related and 0.1 for normal data to handle class imbalance.
Our classifier achieved impressive results on the test set, with an accuracy of 99.5\% and an F1 score of 99.3\%. The true positive rate was 93.02\% and the true negative rate was 99.7\%.

Since the extensions were randomly selected from the entire dataset $D_{t}$, the cross-validation tests generalizability to other unseen extensions. 
With a high true positive rate and a low false negative rate, our approach can effectively uncover 93\% of credential risks, showing its usefulness for systematically identifying potential data exposures.
The robust metrics demonstrate the reliability of our model in discerning credentials from non-sensitive data for new extensions.

\smallsection{False Positives and False Negatives}
The false positive rate is 6.98\% (39/444), where the model failed to identify a data point as credential-related when it actually was. Most false negatives had unusual credential formats that did not match the learned patterns.
For instance, \textit{“haasStudio.wifiSsidPwd”} was one such case where the credential format was not recognized by the model. Additionally, credentials in languages other than English, such as Chinese, posed challenges for the model.
The false negative rate is 0.24\% (39/16,512), where the model misclassified normal data points as credential-related. 
We found two main reasons for these errors: (1) some contain vague keywords like “token” and “key” which are not necessarily credential-related. (2) some data points contained credential-related words but were used for non-sensitive purposes, such as \textit{“toolsHeavenChatGPT.authenticationType”} or \textit{“cloudcode.secrets.createVersionWithText”}. 
Such false positives could be reduced by incorporating more contextual information or using a more fine-grained classification scheme. Nonetheless, the false positive rate is sufficiently low to demonstrate strong performance.

\input{tables/tab3_exposed_type_results}

\subsection{Performance Overhead}
We evaluate the performance of our automated analysis approach on the entire dataset $D_{t}$ consisting of 27,261 extensions.
The experiments are conducted on a machine with a 32-core 64GB RAM, AMD Ryzen 9 5950X 16-Core CPU, and an NVIDIA GeForce RTX 3090 GPU.
The performance overhead focuses on two primary components: (1) processing the source code of extensions and extracting relevant data points from the PDG, and (2) training and evaluating the BERT-based classifier.
For the PDG analysis, we encounter parsing errors for 539 out of 27,261 (2\%) extensions due to the limitations of the Espree parser. Additionally, 11 out of 27,261 (0.04\%) extensions timed out during PDG construction. 
Despite these challenges, we were still able to extract metadata for further analysis. On average, the source code analysis took only 7.37 seconds per extension, amounting to a total of 55.8 hours for all extensions in the dataset.
In the second part, we trained the BERT-based classifier on $D_{gt}$, which took 49 minutes for 5 epochs. The prediction phase for the trained model on $D_{t}$ took 87 hours. The average time to predict one extension was 0.19 seconds.
The performance overhead of our approach is reasonable and acceptable, considering the large scale and diversity of the extensions and the data points. Our approach can efficiently and effectively analyze the VSCode extensions and detect the data exposure risks.

\subsection{Measurement of Impact}

\smallsection{Landscape}
Our study reveals that the issue of potential credential leakage is widespread across the VSCode extension landscape. Out of the 27,261 extensions analyzed, we identified 2,325 extensions that could potentially leak credentials. This represents approximately 8.5\% of the total extensions analyzed, indicating a significant prevalence of this issue.
Furthermore, each of these extensions was found to have an average of 2.11 data items that could potentially lead to credential leakage. This suggests that the risk is not confined to a single data item within an extension, but rather, multiple data items within an extension could potentially contribute to this risk.
These results underscore the critical need for effective measures to detect and mitigate the risk of credential leakage in VSCode extensions. They also highlight the importance of our approach in providing a systematic and efficient means of identifying potential data exposures in a large-scale and diverse dataset of VSCode extensions. This can significantly aid in enhancing the security and privacy of these extensions, thereby protecting the sensitive data of their users.

\smallsection{Exposed Types}
Table~\ref{tab:tab_exposed_type} presents the analysis of different vectors causing credential exposure in VSCode extensions.
Firstly, potential storage access leaks were identified in 1,599 extensions through global state and configuration. Specifically, 316 extensions (18\% of those using global state) leaked credentials via the globalstate API, with each exposing 1.38 unique data items on average. This indicates global state is a common vector for storing and propagating sensitive credentials.
Additionally, requested and used configurations revealed credentials in 1,205 (9.6\%) and 295 (2.7\%) extensions respectively. While lower than globalstate, these configuration leaks still pose worrying risks considering the large user base.
Lastly, in the category of credential control by commands, there are 724 extensions that potentially leak credentials. Within this category, 2.7\% of extensions requested commands and used them to control credential-related data.
These results demonstrate the various ways in which VSCode extensions can expose sensitive data, highlighting the need for effective security mechanisms to prevent such exposures.

\begin{figure}[t]
    \centering
    \includegraphics[width=0.35\textwidth]{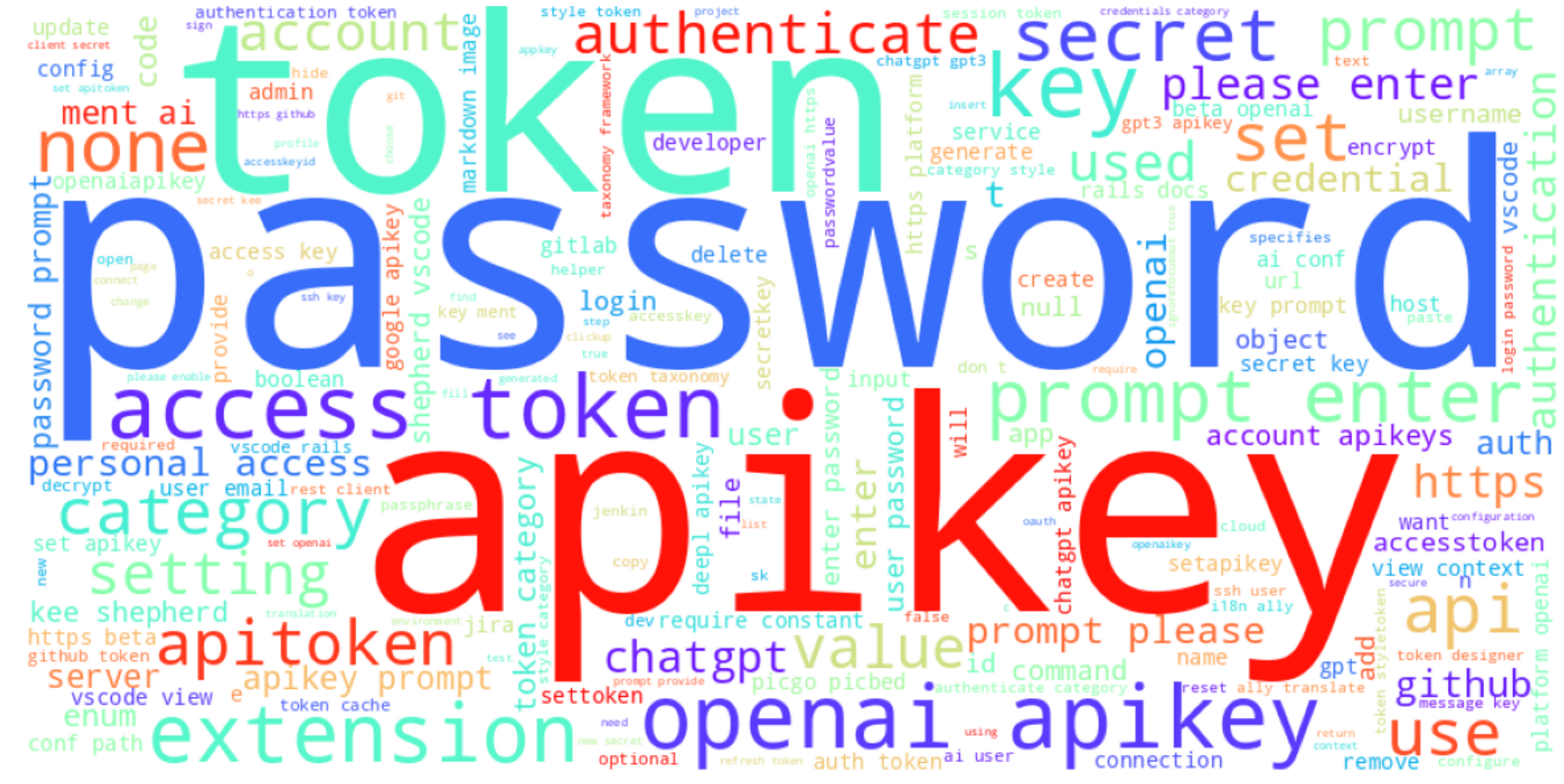}
    \caption{Word Cloud of Exposed Data Items}
    \label{fig:fig_vul_data_wordcloud}
\end{figure}

\smallsection{Exposed Contents}
Figure~\ref{fig:fig_vul_data_wordcloud} presents the word cloud of exposed data items. It is evident that “password” and “apikey” are the most frequent words among the credential-related data, indicating that these are the most common types of credentials that could be leaked by VSCode extensions.
Interestingly, AI-related keywords like “openai” and “chatgpt” also appear in the word cloud. This suggests that as AI-driven extensions like GitHub Copilot~\cite{VSCode_Marketplace_copilot} and Microsoft IntelliCode~\cite{VSCode_Marketplace_IntelliCode} become increasingly popular, the risk of exposing AI-related credentials also rises. This trend underscores the need for heightened security measures in AI-driven extensions, particularly those that require sensitive credentials for operation.
The word cloud also reveals a variety of other potential credential leaks, demonstrating the diversity of data that can be exposed in VSCode extensions.

\smallsection{Data Exposure based on Extension Categories}
To further understand the distribution of data exposure risks across different types of extensions, we grouped the identified 2,325 extensions based on the categories used in the VSCode marketplace. Table~\ref{tab:tab4_vul_over_categories} shows the number and percentage of extensions that potentially leak credentials in each category.
Our findings are surprising, as the percentage of exposed extensions varies significantly between categories.
For example, common categories such as “Keymaps”, “Formatters”, and “Themes” had much lower rates than the overall average of 8.5\%. 
On the other hand, Machine Learning extensions had the highest rate of 40.84\% - almost five times the average. Similarly, Data Science and Education extensions also had high rates of 37.95\% and 23.44\%, respectively.
The high exposure rate in AI/ML domains likely originates from their dependence on external services like OpenAI, Anthropic, and Google, which necessitate authentication. This is particularly alarming considering the growing popularity of AI-based tools.
Moreover, we observed that categories related to security and infrastructure, such as Testing (23.78\%), SCM Providers (28.03\%), and Azure (18.48\%), also had higher-than-average vulnerability rates.
This is likely due to the frequent use of credentials to authorize access to repositories and cloud services.
In conclusion, our results show that data exposure risks are strongly associated with certain VSCode extension categories that involve emerging technologies and security features. As the IDE evolves to support specialized domains like ML and cloud engineering, it is essential to protect credentials and secrets from leakage and prevent serious privacy and infrastructure threats.

\input{tables/tab4_vul_app_over_categories}

\begin{figure}[t]
    \centering
    \includegraphics[width=0.45\textwidth]{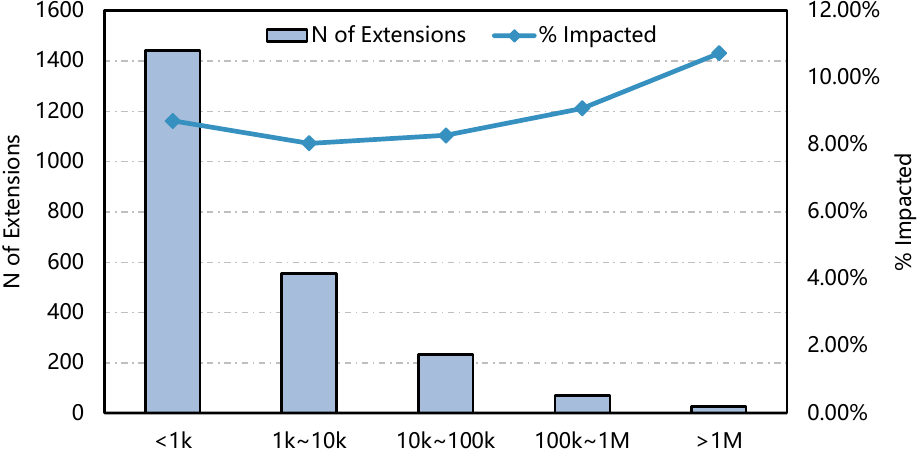}
    \caption{Credential Exposure over Extension Popularity}
    \label{fig:fig_vul_over_installs}
\end{figure}

\smallsection{Data Exposure based on Extension Popularity}
Figure~\ref{fig:fig_vul_over_installs} illustrates the distribution of extensions exposing credentials relative to the number of installations.
Out of the 2,325 extensions detected, more than half (1,441 extensions) have fewer than 1k installations. As the number of installations increases, the number of impacted extensions decreases. However, the percentage of affected extensions gradually increases.
Interestingly, despite only detecting 25 affected extensions with more than 1M installations, the percentage of these extensions is the highest at 10.73\%. This suggests that popular extensions are not immune to data exposure risks, emphasizing the need for robust security measures regardless of an extension’s popularity.

\begin{figure}[t]
    \centering
    \includegraphics[width=0.45\textwidth]{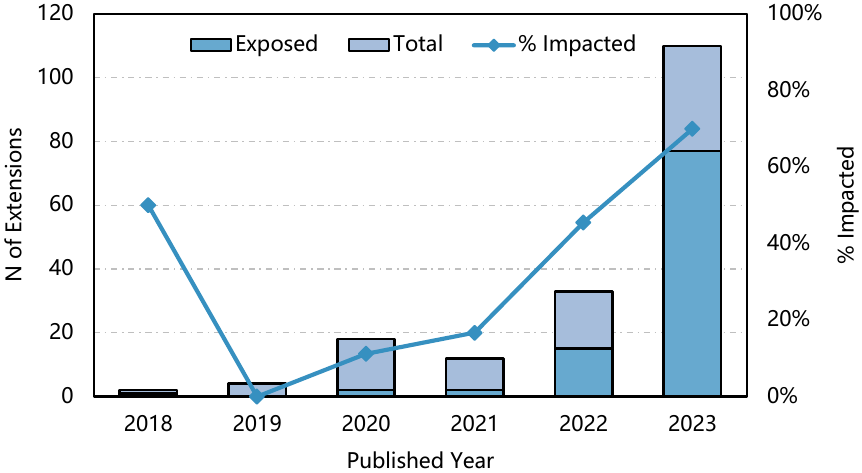}
    \caption{Credential Exposure in AI Coding Assistants}
    \label{fig:fig_vul_aicodingext}
\end{figure}

\smallsection{Data Exposure in AI Coding Assistants}
AI coding assistants are a new trend in the IDE ecosystem, as they aim to help developers write code faster, smarter, and easier. These tools leverage powerful AI models like GPT-4~\cite{achiam2023gpt}, Claude-2~\cite{Anthropic_claude_24}, and Gemini Ultra~\cite{google_gemini_ai} to generate code snippets, suggest code completions, answer coding questions, and even chat with developers~\cite{hou2023large}.
As we described before, however, these AI coding assistants also pose potential data exposure risks, as they may require or leak sensitive credentials for accessing AI services, APIs, or repositories. 
Therefore, in this section, we focus on these AI coding assistants in the VSCode ecosystem.
We used common keywords (i.e., “ai code”, “code completion”, “gpt”, “openai”, “intellicode”, “autocomplete”, “language model”, and “chatbot”) to search the descriptions and tags of the extensions in our crawled dataset $D_{t}$.
We found 179 extensions that fall into the category of AI coding assistants and detected 97 extensions that potentially leak credentials.
This represents 54.2\% of the AI coding extensions, which is much higher than the overall average of 8.5\%.
Figure~\ref{fig:fig_vul_aicodingext} shows the published year of the AI coding assistants and their exposed rate. 
It is obvious that AI coding is increasingly popular, especially from 2022, and only in 2023, there were more than 100 new AI coding assistants published in the marketplace. 
Also, the rate of exposed credentials also increased. For instance, the rate for newly published extensions in 2023 increased to 70\%.
Among the AI coding assistants that could expose credentials, we found that 24 extensions store credentials in GlobalState, including 11 extensions with \textit{“chatgpt-gpt3-apiKey”} and \textit{“chatgpt-session-token”}. Furthermore, 71 AI coding assistants store the credential data in the configuration. This highlights the need for careful handling and secure storage of credentials in these extensions.

%% file: tables/tab3_exposed_type_results.tex
\begin{table}[t]
  \centering
  \caption{Overall Credential Leakage in $D_{t}$}
    \scalebox{0.8}{
    \begin{tabular}{llrrr}
    \toprule
    \multicolumn{2}{l}{\textbf{Exposed Type}} & \multicolumn{1}{l}{\textbf{Items per Exts}} & \textbf{\# Extensions} & \textbf{Total} \\
    \midrule
    \multirow{3}[2]{*}{Storage Access} & GlobalState & 1.38  & 316 (18.0\%) & \multirow{3}[2]{*}{1599} \\
          & Requested Configuration & 1.43  & 1205 (9.6\%) &  \\
          & Used Configuration & 1.23  & 295 (2.7\%) &  \\
    \midrule
    Clipboard Access & InputBox & 1.22  & 620 (11.5\%) & 620 \\
    \midrule
    \multirow{2}[2]{*}{Credential Control} & Requested Commands & 1.65  & 593 (2.7\%) & \multirow{2}[2]{*}{724} \\
          & Used Commands & 1.43  & 458 (2.3\%) &  \\
    \bottomrule
    \end{tabular}%
    }
  \label{tab:addlabel}%
  \label{tab:tab_exposed_type}%
\end{table}%

%% file: tables/tab4_vul_app_over_categories.tex
\begin{table}[t]
  \centering
  \caption{Credential Exposure over Extension Categories}
  \scalebox{0.85}{
    \begin{tabular}{lrr}
    \toprule
    \textbf{Category} & \textbf{N of Extensions} & \textbf{\% Impacted} \\
    \midrule
    Other & 1677  & 8.94\% \\
    Programming Languages & 368   & 8.18\% \\
    Snippets & 225   & 9.07\% \\
    Formatters & 110   & 6.79\% \\
    Linters & 129   & 11.08\% \\
    Debuggers & 75    & 9.18\% \\
    Visualization & 43    & 8.96\% \\
    Extension Packs & 36    & 9.45\% \\
    Themes & 18    & 5.31\% \\
    Testing & 73    & 23.78\% \\
    Keymaps & 11    & 4.98\% \\
    Data Science & 74    & 37.95\% \\
    Education & 45    & 23.44\% \\
    Machine Learning & 78    & 40.84\% \\
    SCM Providers & 44    & 28.03\% \\
    Language Packs & 19    & 15.45\% \\
    Notebooks & 17    & 13.93\% \\
    Azure & 17    & 18.48\% \\
    \bottomrule
    \end{tabular}%
    }
  \label{tab:tab4_vul_over_categories}%
\end{table}%

%% file: sections/discussion.tex
\section{Discussion}
\label{sec:discussion}

\smallsection{Lessions Learned}
The most important lesson learned from our research is the critical need for robust security measures in the development and use of extensions in modern IDEs like VSCode. Our study reveals that even popular and widely-used extensions can have security risks that allow adversaries to stealthily acquire or control sensitive user data. 
This highlights the importance of rigorous security practices in the design and implementation of these extensions.

Another key takeaway is the significant role of cross-extension interactions in the security of VSCode. While these interactions can enhance the functionality and user experience of the IDE, they can also introduce new security risks. Therefore, it is crucial to carefully manage and control these interactions to prevent unauthorized access to sensitive data and resources.

Lastly, our findings highlight the potential risks associated with the increasing use of AI coding assistants in VSCode. These extensions often require sensitive credentials for accessing AI services, which can be exploited by malicious extensions. As such, there is a need for strengthened security measures in AI coding assistants, particularly those that require sensitive credentials for operation.

\smallsection{Mitigation Strategies}
To mitigate data exposure risks in VSCode extensions, we propose strategies for different stakeholders. Extension developers should follow VSCode's security best practices, avoiding plain-text storage of credentials and instead using secure APIs like \textit{SecretStorage}. They should also avoid using command access for credential-related operations. Users should be selective when installing extensions, checking ratings, reviews, and descriptions before installation. They can utilize VSCode's Restricted Mode feature to manage which extensions are enabled for their projects, enhancing their development environment's security. The VSCode platform maintainers should improve the vetting process and security standards for extensions in the marketplace, while also providing additional security features for both developers and users to strengthen the overall VSCode ecosystem security.

\smallsection{Responsible Disclosure}
We have responsibly disclosed our findings to both the VSCode security team and the developers of the affected extensions. We contacted the developers via the email addresses provided on their GitHub repositories or marketplace pages. Out of the 2,325 extensions that potentially leak credentials, we were able to find email contacts for 1,343 extensions. We used a Google Sheets extension called “Yet Another Mail Merge” (YAMM)~\cite{yamm2024} to send personalized emails to each developer and track the status of the emails. Before we submitted this manuscript, the emails were opened by 438 developers, and bounced from 105 invalid email addresses. 
We note that in this process, we only collected the aggregated results of the emails provided by YAMM, and we never inspected the tracking information of individual addresses. Moreover, we did not obtain any information about or from individuals.
Among the 438 opened emails, 15 developers acknowledged our findings and promised to fix them in future versions.
Two developers also reported false positives due to semantic errors. 

Due to the low response rate from these extension vendors, we also reported these risks to Microsoft as a product security issue, since they provide security guidelines for extension development. They acknowledged our findings and assured us that they had shared the report with the team responsible for maintaining the product or service.

\section{Limitations}
Our work has several limitations that we acknowledge and discuss in this section.
First, false positives and negatives can arise for a few reasons.
Our analysis is based on the source code and metadata of VSCode extensions, which may not reflect the actual runtime behavior of the extensions. 
For example, some extensions may use code obfuscation techniques to protect their sensitive data, which we cannot detect by static analysis. 
Further, our approach relies on a fine-tuned BERT model to classify the data points as credential-related or not.
Although our model achieves high accuracy and F1 score, it may make mistakes in some cases. Our model may fail to recognize some uncommon credential formats or misclassify normal data points as credential-related.
Our ground truth data also focused narrowly on common secrets like API keys and passwords, potentially overlooking other sensitive data types. 
These factors contribute to both false positives and false negatives.

Second, our approach does not consider the dynamic and evolving nature of the VSCode extension ecosystem. 
As extensions continue to advance, new exposure vectors may arise outside of our analyzed data set.
Moreover, while discovering many potentially vulnerable extensions, we lack measurements of real-world exploitability and user impact. 
Quantifying exposure severity could better demonstrate security implications for developers and users. Furthermore, our findings are restricted to VSCode extensions, whereas other IDEs utilizing extensions may introduce distinct threats. Comparative research across IDE ecosystems could reveal further dimensions.

%% file: sections/related_work.tex
\section{Related work}
\label{sec:relatedwork}

\smallsection{Software Developer Tools Security}
Software developer tools, including IDEs, code editors, and compilers, are crucial for software development but can introduce security and privacy risks~\cite{lin2024untrustide, pearce_asleep_2022, liu2019metaheuristic}. Package management systems like npm and PyPi, while facilitating code reuse, pose challenges such as dependency confusion and malicious code injection. Zimmermann~\ea~\cite{zimmermann2019small} and Alfadel~\ea~\cite{alfadel2023empirical} conducted large-scale studies on these ecosystems, proposing mitigation strategies. Code generation tools, such as IDE plugins, can also introduce vulnerabilities like injection attacks and broken access control. Li~\ea~\cite{li2019evaluation} evaluated IDE plugins, highlighting common security issues. Pearce~\ea~\cite{pearce_asleep_2022} analyzed GitHub Copilot, finding that 40\% of its generated code contained security vulnerabilities. Cloud-based development platforms, such as Amazon EC2 and Microsoft Azure, expose developers to risks like co-location attacks. Varadarajan~\ea~\cite{varadarajan2015placement} proposed secure placement policies to mitigate these risks.

\smallsection{Privacy Leakage Detection}
Privacy leakage detection aims to prevent the exposure of sensitive information from software systems using techniques like static analysis, dynamic analysis, and machine learning~\cite{nan2015uipicker, nan2018finding, zuo2019does, nan2023you, meli2019bad, zhang2023don, feng2022automated, chen2018mystique, liu2022deep}. UIPicker~\cite{nan2015uipicker} and ClueFinder~\cite{nan2018finding} identify sensitive data from app UIs and user input using semantic features and program structures. Zuo~\ea~\cite{zuo2019does} explored cloud data leaks, attributing them to authentication and authorization misconfigurations, and developed LeakScope to detect such vulnerabilities. Feng~\ea~\cite{feng2022automated} introduced PassFinder for detecting password leaks in public repositories. Chen~\ea~\cite{chen2018mystique} created a taint analysis framework for browser extensions, revealing significant privacy risks. These studies underscore the need for effective privacy leakage detection tools to safeguard sensitive information across various software systems. This revised section maintains the essential information while being more concise and aligned with the manuscript's style.

%% file: sections/conclusion.tex
\section{Conclusion}
\label{sec:conclusion}
In this paper, we have identified and demonstrated a new category of security risks posed by cross-extension interactions in Visual Studio Code. 
This study, the first of its kind, reveals that malicious extensions can exploit the security flaws of other extensions or VSCode itself to steal or manipulate sensitive data (e.g., passwords, API keys, access tokens, etc.). 
% We demonstrate the significant impacts of these attacks by developing and publishing our own proof-of-concept extensions in the official marketplace.
We designed an automated analysis framework leveraging program analysis and language models to systematically detect such risks at scale.
By applying our tool to over 27,000 real-world VSCode extensions, we discovered over 2,325 extensions that improperly expose credential-related data through commands, user input, and global state.
Our study brings new insights and implications for the security of VSCode extensions and the extension-in-IDE paradigm.

\section*{acknowledgment} 

The authors would like to thank the anonymous reviewers who have provided insightful and constructive comments on this paper. 
Chakkrit Tantithamthavorn was partly supported by the Australian Research Council's Discovery Early Career Researcher Award (DECRA) funding scheme (DE200100941).